\documentclass[aip,
 amsmath,amssymb,
 reprint,%
]{revtex4-2}

\usepackage{graphicx}
\graphicspath{{./figures/}{./figures_supp/}}
\usepackage{dcolumn}
\usepackage{bm}

\usepackage[utf8]{inputenc}
\usepackage[T1]{fontenc}
\usepackage{mathptmx}
\usepackage{etoolbox}
\usepackage{hyperref}
\usepackage{slashed}
\usepackage[normalem]{ulem}

\usepackage{color}
\definecolor{gold}{rgb}{0.83, 0.69, 0.22}
\definecolor{green}{rgb}{0.0,0.5, 0.0}

\newcommand{\SG}[1]{\textcolor{magenta}{{ #1}}}
\newcommand{\SGs}[1]{\textcolor{magenta}{{\sout{#1}}}}
\newcommand{\CP}[1]{\textcolor{green}{{#1}}}

\newcommand{\ba}[1]{{\bf a}}

\newcommand{\bs}{\boldsymbol}

\makeatletter
\def\@email#1#2{%
 \endgroup
 \patchcmd{\titleblock@produce}
  {\frontmatter@RRAPformat}
  {\frontmatter@RRAPformat{\produce@RRAP{*#1\href{mailto:#2}{#2}}}\frontmatter@RRAPformat}
  {}{}
}%
\makeatother
\begin{document}

\title{High temperature melting of dense molecular hydrogen from machine-learning interatomic potentials trained on quantum Monte Carlo}
\author{Shubhang Goswami}
 \affiliation{The Grainger College of Engineering, University of Illinois Urbana-Champaign, Urbana, Illinois 61801, USA}
\author{Scott Jensen}
\affiliation{The Grainger College of Engineering, University of Illinois Urbana-Champaign, Urbana, Illinois 61801, USA}
\author{Yubo Yang}
\affiliation{Center for Computational Quantum Physics, Flatiron Institute, New York, NY 10010, USA}
\affiliation{Department of Physics and Astronomy, Hofstra University, Hempstead, NY 11549, USA}
\author{Markus Holzmann}
\affiliation{Univ. Grenoble Alpes, CNRS, LPMMC, 38000 Grenoble, France}
\author{Carlo Pierleoni}
\affiliation{Department of Physical and Chemical Sciences, University of L'Aquila, Via Vetoio 10, I-67010 L'Aquila, Italy}
\author{David M. Ceperley}
\affiliation{The Grainger College of Engineering, University of Illinois Urbana-Champaign, Urbana, Illinois 61801, USA}

\date{\today}

\begin{abstract}

We present results and discuss methods for computing the melting temperature of dense molecular hydrogen using a machine learned model trained on quantum Monte Carlo data. In this newly trained model, we emphasize the importance of accurate total energies in the training. We integrate a two phase method for estimating the melting temperature with estimates from the Clausius-Clapeyron relation to provide a more accurate melting curve from the model.  We make detailed predictions of the melting temperature, solid and liquid volumes, latent heat and internal energy from 50 GPa to 180 GPa for both classical hydrogen and quantum hydrogen. At pressures of roughly 173 GPa and 1635K, we observe molecular dissociation in the liquid phase. We compare with previous simulations and experimental measurements.

\end{abstract}

\maketitle

\section{Introduction}

Hydrogen is the first element in the periodic table. Under extreme conditions, it  has many important applications in diverse fields such as in astrophysics, and for inertially confined fusion. Understanding its phase diagram has been a long standing quest for both theory\cite{RMP2012} and experiment since the pioneering metallization predictions of Wigner and Huntington in 1935\cite{Wigner1935}. However, definitive prediction of its properties at high pressure has presented difficult computational and experimental challenges. Experimentally, it has proved difficult to confine hydrogen in diamond-anvil cells for temperatures above 300K and pressures above 300GPa so that less precise dynamical methods are employed.  
On the computational side, predictions using density functional theory (DFT) are uncertain because of the approximate nature of the exchange-correlation functional, the importance of dispersion effects and the electronic gap closure during the atomic-molecular transition. Molecular hydrogen poses other difficulties because of the large quantum effects of the protons and disorientation of the molecular bond vectors in the crystalline phase I. 

A major focus of recent theoretical and experimental effort has been on verifying the atomic-molecular transition, also known as the liquid-liquid phase transition (LLPT). Our recent simulations\cite{Niu2023}, using a more accurate intermolecular potential with Quantum Monte Carlo (QMC) accuracy, found a much higher melting temperature than was seen in earlier DFT simulations and hinted at in experiments. The increased stability of the solid would imply that the LLPT may be largely hidden within the crystalline molecular phase and not seen in the liquid phase. In this paper, we develop a substantially improved machine-learned interatomic potential (MLIP) by putting more weight on the description of the energy than on the forces. Using this MLIP we compute more accurate melting temperatures of molecular hydrogen and properties along the melting curve.

In the 60 years since A. Rahman's work\cite{Rahman1964} on liquid argon, the field of molecular dynamics (MD) has blossomed. In this work, we will use the Parrinello-Rahman constant pressure MD\cite{Parrinello1980} to estimate the melting temperature of molecular hydrogen at pressures sufficiently high that experimental data is lacking, difficult to interpret or unreliable.  We also use the path integral molecular dynamics approach developed by Parrinello and Rahman\cite{Parrinello1984} to include effects of the quantum motion of the protons. The challenge Rahman had was to come up with accurate interaction energies and forces (the potential energy surface (PES)) to drive the dynamics.  
More realistic PES were obtained using the ``ab initio'' approach of Car and Parrinello\cite{CarParrinello1985} where a density functional solution of the electronic wavefunction is used to obtain the PES.  This achieved much higher accuracy than simple model potentials, but accurate DFT functionals are still difficult to find, particularly for systems undergoing phase changes and those that lack clean experimental measurements to benchmark the functionals.  Finding accurate interactions for the various phases of dense hydrogen presents particular challenges because hydrogen molecules have important dispersion interactions while atomic hydrogen has long-range Coulomb interactions.  The transition between the two states is a metal-insulator transition with the electronic gap closing. This transition is difficult for common functionals to get correct. Finally, the protons have a large, anharmonic quantum motion. 

Quantum Monte Carlo (QMC) can provide a more accurate energy surface (PES) than from the usual DFT functionals such as PBE or vdw-DF as we discuss below. \footnote{Our calculations, as well as all others that we are aware of, start with the Born-Oppenheimer approximation (BOA). It assumes that the electrons are in their ground state for a given arrangement of protons and can be integrated out. We are left with the problem of finding the equilibrium properties of the protons interacting with a PES. We refer to classical hydrogen as the system having classical protons and quantum hydrogen when they have their physical mass. In earlier work we estimated the effect of the BOA on the energy at 10 K per atom at low density. The correction of the energy of liquid and solid will be very similar, so it will have a negligible effect on the melting line.} 
Although computational resources needed for QMC are several orders of magnitude higher than for DFT, hydrogen is a good case for its use since it has only a single electron per atom, one does not need to use a pseudopotential thus eliminating this approximation and reducing the computational effort, and very accurate trial functions are available\cite{PIERLEONI2008}.

Over the past 20 years we have developed a purely QMC procedure to simulate dense hydrogen, the Coupled-Electron Ion Monte Carlo (CEIMC) method. Using this method we have performed simulations of the liquid-liquid transition\cite{Morales2010}, investigated the properties of lower temperature crystalline hydrogen\cite{Rillo2018} and determined optical properties of dense hydrogen\cite{Gorelov2020}. However, when we attempted to use this procedure to determine the melting transition of phase I of molecular hydrogen, we were surprised to find a higher melting temperature than reported experimentally. This initial investigation
was performed using a simulation with both a vdW-DF1\cite{vdw-df} density functional and a QMC energy surface with supercells of 48 molecules. 
Based on the Bragg peaks, melting occurred between 1750K and 2000K \CP{\footnote{see Supplementary Material section in ref \cite{Niu2023}}}. Such estimates of the melting temperature were not very precise because of errors due to finite size effects and incomplete convergence of the trajectories. Despite these uncertainties, the results raised questions about the estimated melting temperature of molecular hydrogen for pressures above 45 GPa. As we discuss below, these estimates are based on using the PBE density functional in conjunction with dynamical experiments that provide very limited information on the state of the experimental sample.   

In order to overcome the limitations of CEIMC we constructed a database containing 16290 configurations of dense hydrogen with QMC forces and energies: see ref. \cite{Niu2023} for details. In ref.~\cite{Niu2023} we used this database to train a neural network potential with Deep Potential Molecular Dynamics (DPMD)~\cite{Zhang2018,Zhang2018a} to describe dense molecular hydrogen.
In this paper, we present a new estimate of the high pressure melting line of molecular hydrogen using a different machine-learned (ML) model, MACE~\cite{Batatia2022mace, Batatia2022design}, which we believe is a better representation of the QMC data and hence of actual hydrogen. We use this new model to make more precise calculations of the high pressure molecular hydrogen melting line.

\section{ Machine learning model}
Our previously developed hydrogen database\cite{Niu2023} was used to train a MACE model~\cite{Batatia2022mace, Batatia2022design} designed to address high pressure molecular hydrogen. However, as molecular hydrogen is compressed along the melting curve from ambient pressure at low temperature to 180 GPa at 1800K, the molar volume decreases from 23.1 cc/mol to 2.4 cc/mol: a factor of ten. 
This large compression has 
consequences in developing a ML model since  
the ML model has a cutoff distance in constructing the descriptors of a configuration. To describe the low pressure molecular solid, the cutoff needs to include at least the 12 nearest molecules in the hcp lattice.  Using the same model at 180 GPa, there would be 10 times the number of molecules within the cutoff or about 240 atoms, an 
inconveniently large number of atoms. Our QMC calculations to determine the forces and energies are done on systems of 96 atoms, a significantly smaller number. To solve this problem we restricted the training to configurations in the range 50GPa $\leq$ P $\leq$ 200 GPa. Over this range the volume only changes by a factor of 1.7.
We restricted the range of temperature as well so as to primarily train on liquid and solid molecular hydrogen, i.e. under conditions having a small number of dissociated molecules.

The 80,000 configurations in the database were sampled from a grid in pressure (50 $\leq$ P $\leq$ 200GPa) and temperature (300 $\leq$ T $leq$ 2000K) using ab-initio molecular dynamics (AIMD) for classical hydrogen and Path Integral Molecular Dynamics (PIMD) for quantum hydrogen.
A subset of 16,290 configurations were selected for QMC calculations of forces and energies. 
The average statistical errors of the QMC energies and forces are shown in Table \ref{tab:hydrogen errors}.  See the Appendix  and Niu et al.\cite{Niu2023} for further details.

\begin{table}[]
    \centering
    \begin{tabular}{|c|r |r |}
    \hline
     PES  & $\epsilon_E$ & $\epsilon_F$  \\\hline
       QMC  & 1.1 &  77  \\\hline
       PBE   &336 &    418 \\
       vdW   &124 &    153  \\\hline
       DPMD &127   &   287 \\
       MACE   &27 & 133 \\
        \hline
    \end{tabular}
    \caption{Root mean squared errors in the energy (in meV) and forces (in meV/\AA) averaged over 16,290 configurations each with N=96 protons.
    Because the total energy error in a disordered system is the sum of local uncorrelated errors, the energy variance is extensive in the system size. To allow fair comparison with different values of N, we divided the error in the total energy by N$^{1/2}$. The first row is the statistical error of the QMC data. The next two rows are the errors of DFT functionals with respect to the QMC results. The last two rows show the errors in the DPMD model\cite{Niu2023} and MACE M18 model trained on QMC data with respect to the QMC results.
     }
    \label{tab:hydrogen errors}
\end{table}


Our previous DPMD model
focused on reproducing the forces from QMC rather than the total energy~\cite{Niu2023,Ceperley_2024}.
In preparing a new model for the current work, we made a study of increasing the importance of the energy errors versus the force errors in training the new model. 
The ``loss function'' that is minimized to determine the ML model is:
\begin{equation}
\mathcal{L}=\frac{\lambda_E}{BN^2}\sum_{b}^{B}(\hat{E}_b-E_b)^2+\frac{\lambda_F}{3BN}\sum_{b,I,\alpha}^{B, N,3} \left( -\frac{\partial \hat{E}_b}{\partial R_{b,I,\alpha}}-F_{b,I,\alpha} \right) ^2
\label{eq:loss}
\end{equation}
where $B$ is the number of configurations $R_b$, $N$ the number of atoms, $E$ and $F$ the QMC estimated energies and forces respectively.$\hat{E}_b$ is the ML energy for configuration $b$. $\lambda_E$ and $\lambda_F$ are parameters controlling the importance of the energies and forces to the fit. The QMC data has 96 atoms in a supercell, so there is much more data describing the forces than the energies: 288 times as much.  Although the force data is crucial since it is local, it tends to overwhelm the information in the total energy.

It is important to have small energy errors since the relative stability of two phases is determined by energies. In this work, we estimate the melting temperature of molecular hydrogen. The difference in the Gibbs free energy of the liquid and solid phases determines the melting conditions. A major component of that is the difference in their internal energies.  
Although the training data on the forces might describe each phase accurately, unless the ML model gets the energy difference accurately, the melting temperature will be biased. In order for force training to get that difference correctly, the training data must have accurate forces in the transition region between liquid and solid since it is the integral of the forces across the transition region that will determine the relative energies. Since that region has a higher energy it will be infrequently sampled. Training with energies circumvents this problem. 
In addition to this thermodynamic argument, QMC energies are more accurate than QMC forces: the QMC energy error is second order in nodal errors versus first order for the forces and the energy does not have a mixed estimator bias. Even though for a perfect model, energies and forces would be equivalent, when one fits an imperfect model with noisy data, the accuracy of the results can differ. 

\begin{figure}
    \centering
    \includegraphics[width=0.9\linewidth]{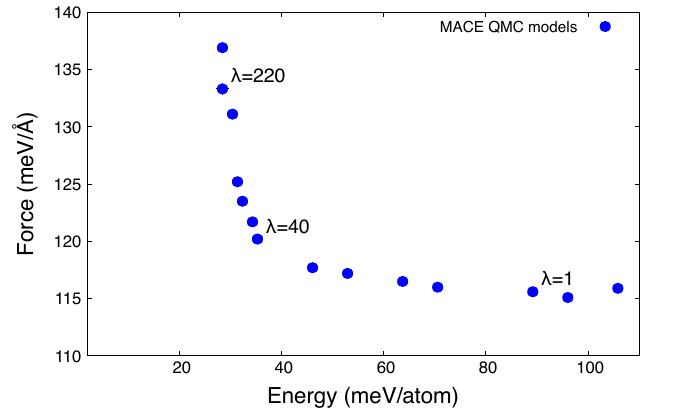}
    \caption{RMS force errors versus the RMS energy errors as $\lambda=\lambda_E/\lambda_F$ varied. The M18 model used in this paper  has $\lambda=220$\AA $^{-2}$. Errors are the two terms in the loss function of Eq. (\ref{eq:loss}) averaged over the 16290 configurations in the database.
    }
    \label{fig:MACE_Pareto}
\end{figure}

In this work, we trained MACE models
with increasing complexity of architectures and larger atomic environment cutoff radii until the errors saturated keeping in mind the MD run times of a model.  We then increased the energy weight until the energy error saturated, see Fig. \ref{fig:MACE_Pareto}, to obtain the final model which we will call the MACE-M18 model. 
Compared to our previous DPMD model, the energy errors are reduced by a factor of 4.5 and the force errors by a factor of 2.6. See Table \ref{tab:hydrogen errors}. These errors are averaged over all QMC energies and forces in the data base with the QMC stochastic errors accounted for\cite{Ceperley_2024}. The MACE model is fitted directly to the QMC data, while the DPMD used a hierarchical approach\cite{Niu2023}, first by subtracting out the molecular binding energy, then by subtracting a model trained on PBE energies and forces. Although the M18 model has much smaller energy errors than the DPMD model, the QMC stochastic error is very much smaller, suggesting a considerable room for further improvement in ML models of hydrogen.  

In addition to monitoring the difference between the model and the QMC results, we have done a number of comparisons of the M18 model with previous experimental and computational results. Among the tests are:
\begin{itemize}
\item Shown in fig.~\ref{fig:MACE_V_ctoa_vs_P} is the comparison of the density versus pressure (the equation of state) of solid hydrogen with experimental x-ray scattering data at room temperature. The c/a ratio of the M18 model is in closer agreement with the experimental data than is the DPMD model. Note that there is a transition from Phase I to Phase III at 200GPa, which could be difficult to get precisely correct. 

\begin{figure}
    \centering
    \includegraphics[width=0.8\linewidth]{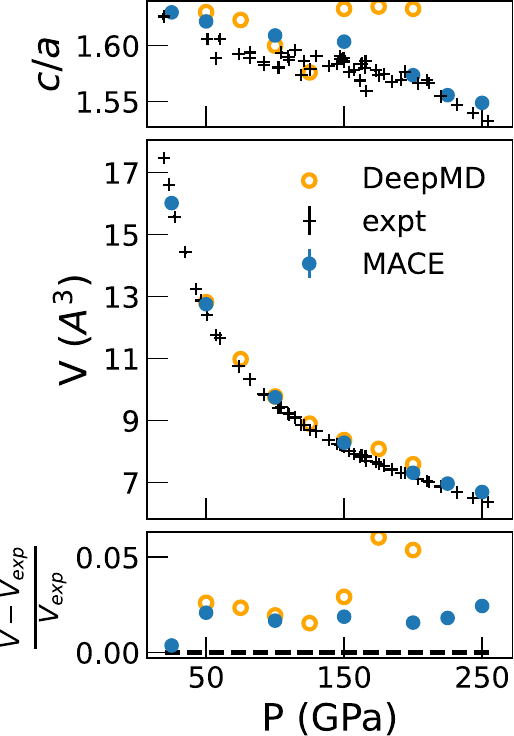}
    \caption{A comparison of the c/a ratio (top panel), equation of state (middle panel) between our two machine learned models and experimental X-Ray scattering measurements\cite{Ji2019} at room temperature as a function of pressure. The bottom panel shows the relative deviation between the simulation results and the experimental atomic volumes.
    Results from the MACE M18 model are in closer agreement with the experiment than those from the previous DPMD model~\cite{Niu2023} at 200 GPa. 
    }
    \label{fig:MACE_V_ctoa_vs_P}
\end{figure}

\item Shown in fig.~\ref{fig:BOPIMC_vs_MACE_gr} is a comparison of the pair correlation function with CEIMC calculations in the molecular liquid near the melting line at 1500K. CEIMC calculations use Path Integral Monte Carlo (PIMC) for the protons and variational MC for the electron degrees of freedom. Hence the CEIMC results do not have a machine learning approximation.  

\begin{figure}
    \centering
    \includegraphics[width=0.8\linewidth]{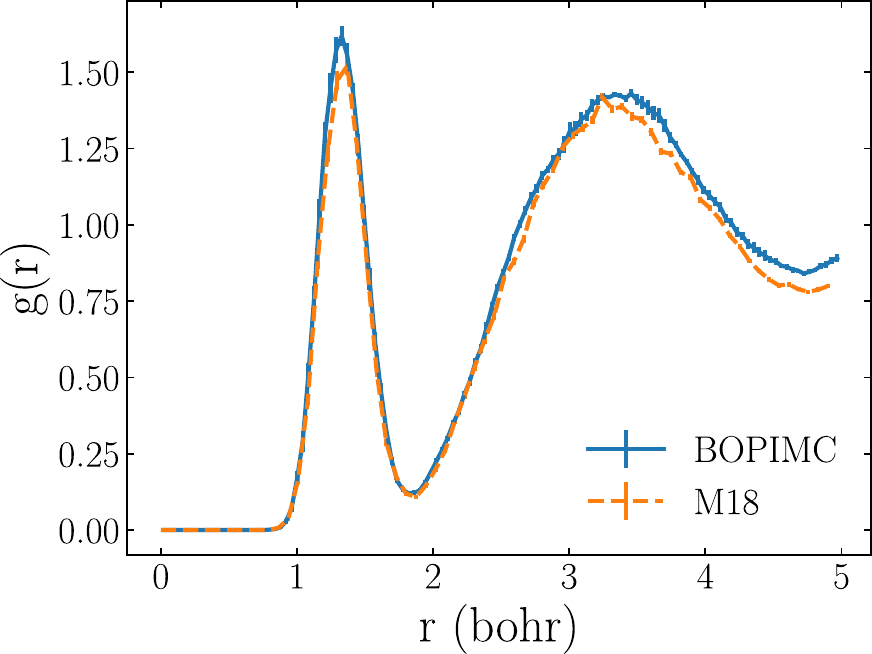}
    \caption{A comparison of the pair correlation function of CEIMC calculations\cite{Pierleoni2016b} and the MACE M18 model for the molecular liquid at 1500 K and $r_s=1.55$ bohr (v=2.784 cc/mol). The M18 simulation was performed using 96 protons, whereas the CEIMC calculations were performed using 54 protons. Both included ZPE of protons using path integral MD or MC, respectively.} 
    \label{fig:BOPIMC_vs_MACE_gr}
\end{figure}

\item Using DPMD trained on PBE agrees with previous PBE ab initio simulations of the melting transition of molecular hydrogen. See Niu et al.\cite{Niu2023} for details. This shows the model can accurately reproduce the results of a realistic simulation of molecular hydrogen.
\end{itemize}

\section{ Melting methods}
Here we discuss various methods to determine the melting temperature.
By comparing various methods we can estimate biases and errors.
There have been extensive comparisons of algorithms that determine melting temperatures\cite{ZhangMaginn2012,WANG2024}. We would like more reliable estimates of the melting temperature to enable detailed comparisons with other models and with experiment.

The simplest approach to determine the melting transition is to start from a liquid or solid structure and slowly raise or lower the temperature, until the system changes into the other phase.  This will give upper and lower bounds; however such bounds are not very tight. 

The most accurate approach is to make a calculation of the free energy using thermodynamic integration. 
In the solid, one can turn off the ML interactions and turn on an intramolecular pair potential and a harmonic interaction to bind the center of mass of each molecule to a single lattice site.
In the fluid, again we need to turn off the ML potential and turn on an intramolecular potential to reach the ideal gas limit of diatomic molecules at low density or high temperature. 
 Morales et al.\cite{Morales2010} and Liberatore et al.\cite{Libertore2011} describe this approach to determine the melting transition in dense hydrogen.
 However, this method requires careful control of numerical errors so that neither the liquid nor the solid are biased and have finite size errors that will cancel.  A path through phase space to a reference point with known free energy must be found that does not cross a phase boundary.

There are several features in the melting transition of molecular hydrogen that differ from other systems.  First, in both the liquid and solid phase the bond angles are disoriented along the melting line. The classical system will have oriented bonds at low temperatures but quantum effects restore the disordered state. Second, we wish for a method that computes the entire melting curve over a broad range of pressure. Finally, the volume change at melting is small, less than 2\%, implying that the hydrogen system can readily melt or freeze. Hysteresis, typically seen with melting, is almost absent.
A recent work on atomic hydrogen\cite{ly2024} computed melting with similar methods to those used here. However, atomic hydrogen is simpler lacking the molecular-atomic transition and bond dynamics. In that work, we did not compare various methods for determining melting temperatures.

To begin, we assume classical protons. While quantum effects of the protons are significant, they are less so at elevated temperatures. It is useful to test the various methods on classical systems and, as a second step, incorporate quantum effects based on the classical melting curve. 

\subsection{Two phase (TP) method}

In the two-phase (TP) method to determine melting, an initial proton configuration consisting of half hcp solid and half liquid is constructed.  The construction of this initial condition is quite important; if the interface between the phases has excess energy it can be difficult for the thermostat and barostat to guide the system to quasi-equilibrium with both phases still present in a reasonable number of steps. 
In our simulations, the initial solid had 32 total layers with 24 molecules in each layer aligned so their basal plane was perpendicular to the z-axis.
A liquid configuration of the same size was constructed by melting the solid configuration and then quenching it to the same temperature as the solid. The liquid and solid configurations were then placed in a common cell with the solid occupying $z<0$ and the liquid $z>0$   
Periodic boundary conditions were then used in all 3 dimensions so that the initial cell has two liquid-solid interfaces. 
With this number of layers, 1/16 of the solid and 1/16 of the liquid started on a layer adjacent to the other phase. If the surface remains planar, this fraction remains constant. However, roughening of the surface increases the area of the interface. The system's evolution under MD at constant temperature and pressure was monitored to determine whether it melted, froze or remained in a mixed phase.

\begin{figure}
    \centering
    \includegraphics[width=0.9\linewidth]{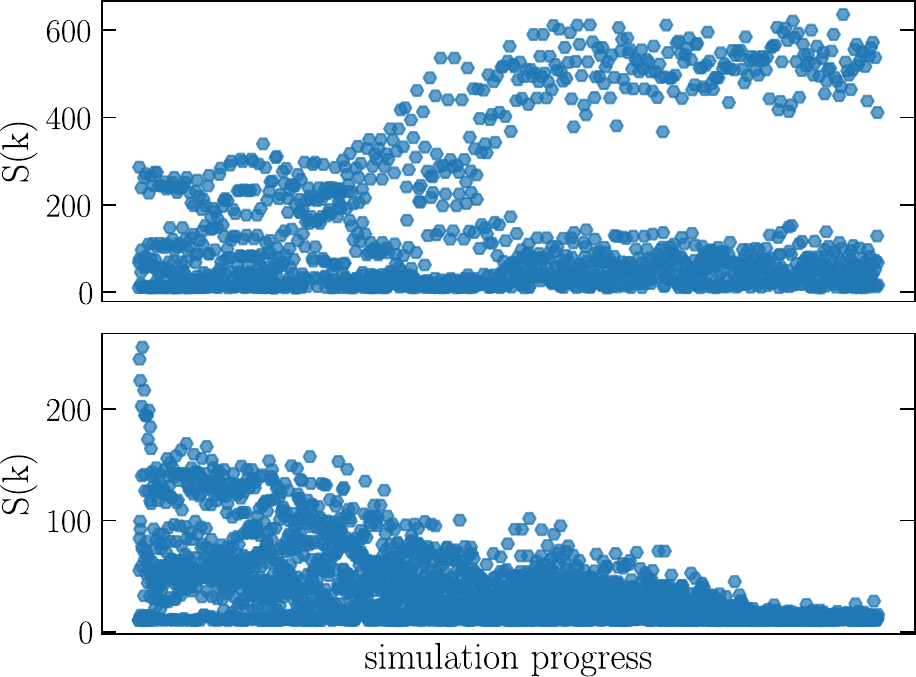}
    \caption{The structure factor S$({\bf k})$ for wavelengths with 4.15 \r{A}$^{-1} \leq k \leq $ 4.75 \r{A}$^{-1}$(i.e. around the h.c.p. Bragg peaks) for the two phase simulation with 3072 atoms at 100 GPa as a function of simulation time. The top panel (1000 K) shows an example of the two phase system freezing; the bottom panel (1400K)  shows the system melting.
    }
    \label{fig:braggpeaks}
\end{figure}
We used two correlation functions to monitor the presence or absence of crystalline order. 
First, a strongly layered structure of the atomic density profile along the z-direction indicates crystalline order where the number of these defined layers is a measure of the extent of the solid. See Niu et al.\cite{Niu2023} for an illustration.
Second, we computed the instantaneous structure factor for wavelengths associated with ordering and monitored its magnitudes during the simulation.  Examples of melting and freezing are shown in fig. ~\ref{fig:braggpeaks}.

At high temperatures, the whole system melts, giving an upper bound to the melting temperature. Conversely,  at low temperatures, the system freezes, giving a lower bound. Having established upper and lower bounds, we adjust the temperature to tighten the bounds. We have established bounds for classical hydrogen at six different pressures from 50 GPa to 175 GPa as shown in fig \ref{fig:melt}. The tightest bounds were 25K, while the loosest were 100K.

Thermal fluctuations can cause the system to end up in the incorrect phase for systems with 1536 molecules.  If the temperature and pressure are exactly on the melting curve, the dynamics of the location of the liquid-solid interface can be thought of as a one-dimensional random walk and the termination of the process with complete melting or freezing as the classic problem of ``first passage time''  of a random walk.
Keep in mind that if the number of liquid or solid layers is fewer than about 3, then the minority phase will not be thermodynamically stable due to surface energies. If the temperature is slightly above or below the melting point, there will be a drift proportional to $T-T_m$, pushing the system toward the correct phase. Only when this drift dominates the fluctuations can we be certain that we have found upper or lower bounds.

Given this argument, we demand that the established bounds be repeatable, independent of initial conditions, and that the drift dominates over the fluctuations. However, these simulations are very slow, especially to make tight bounds because the drift vanishes as the transition is approached. The supercell contains many molecules and hence requires large computer resources.  Another procedure is needed to improve the accuracy, reliability and efficiency of the estimation of the melting curve. 

\subsection{Using Clausius-Clapeyron (CC) relation with the two phase method}  

The Clausius-Clapeyron (CC) relation links the derivative of the melting temperature with respect to pressure to quantities calculable within simulations of homogeneous systems\SGs{:}\SG{,} namely the specific volumes and internal energies of the liquid and solid phases. Simulations of homogeneous systems are less computationally demanding, both in terms of number of atoms and the number of iterations required for convergence.

Kofke \cite{Kofke1993} developed a method to use the CC relation to determine the entire liquid-gas coexistence curve using a sequence of simulations along the coexistence curve at increasing temperatures up to the critical point. While we use the CC relation for a sequence of pressures, we perform a statistical analysis to find possible melting curves consistent with our knowledge from both the TP bounds and the homogeneous CC data.

At various pressures ranging from 50 GPa to 180 GPa (see Table \ref{table1}) we performed constant (P,T) MD runs scanning the temperature above and below a suspected melting temperature as established by the TP  method described above. Solid runs were seeded with a perfect hcp crystal configuration, with all of the molecular bonds pointing out of the basal plane. A barostat allowed for the overall volume and the c/a ratio to fluctuate and to adjust dynamically to a stress-free equilibrium.  Liquid runs were seeded by quenching a well-equilibrated 2000 K configuration with the simulation cell constrained to be cubic. We also monitored to ensure the system remained in a molecular phase when preparing our liquid configurations and throughout our simulations. 

We checked that neither the liquid nor the crystal fluctuated into the other phase by monitoring the energies, volumes and the structure factor; such fluctuations are quite common. We found that the energies and volumes of 192 atoms in the simulation cell agreed with those of 360 atoms. Since the MACE model is a short-ranged potential we do not expect an unusual size dependence.  

\begin{figure}
    \centering
    \includegraphics[width=0.97\linewidth]{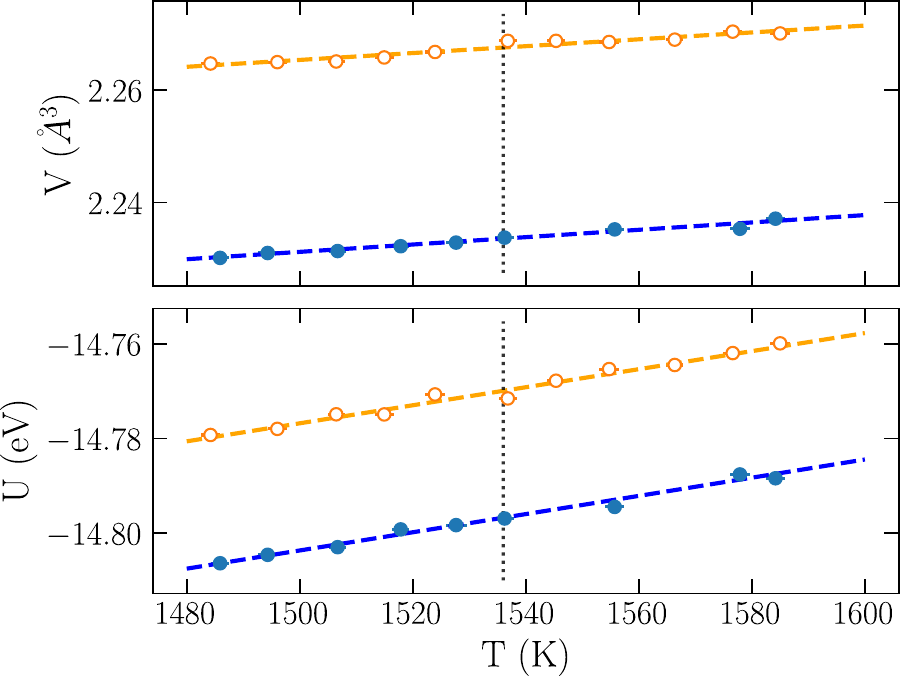}
    \caption{Example of the classical MD results showing volume per atoms in \AA$^3$ (top) or energy per atom in eV (bottom) versus temperature at 130 GPa for 360 atoms. Upper orange open symbols and lines represent data from simulations starting from a liquid configuration, while lower blue lines with closed symbols represent data from simulations starting with an hcp configuration. The energy and volume differences are used to determine the CC derivative at this pressure. The vertical dotted line indicates the estimated melting temperature.
    }
    \label{fig:n130}
\end{figure}

Using these temperature scans for each pressure, we fit the liquid and solid energies and volumes to linear relations in temperature allowing us to interpolate to intermediate temperatures; see the example shown in fig. \ref{fig:n130}. 
In order to use the CC relation we first construct the following function of temperature for a given pressure, $P$:
\begin{equation}
    D(T,P) =T / \left( P + \frac{u_l-u_s}{v_l-v_s }   \right)
    \label{eq:D}
\end{equation}
where $u$ is the internal energy per atom, $v$ the volume per atom (both at temperature T and pressure P);  the subscripts $\{l,s\}$ refer to the liquid and solid phases. 
We compute the statistical error, $\sigma (P)$, in $D(P,T)$ by propagating the errors in the energies and volumes.
For $T_m$  on the melting line, the CC relation gives: 
\begin{equation}
    \frac{dT_m(P)}{dP}=D(T_m,P) .
    \label{eq:cc}
\end{equation}


\begin{figure}
    \centering
    \includegraphics[width=\linewidth]{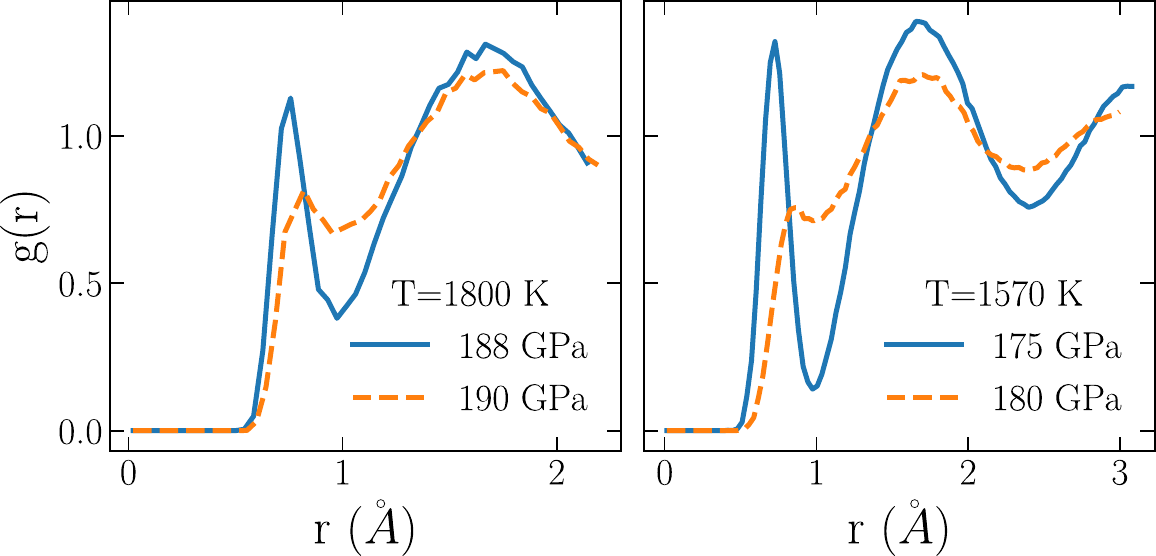}
    \caption{ Pair correlation functions for classical hydrogen at 1800 K at pressures 188 GPa and 190 GPa in the liquid phase (left) and quantum hydrogen at 1570 K at pressures 175 GPa and 180 GPa in the liquid phase (right). The peak at the H$_2$ bond length of 0.75\AA \  is proportional to the number of molecules. Hence, most molecules dissociate abruptly over these small pressure increases. }
    \label{fig:gofr_triplepoint}
\end{figure}

We assume that the melting curve (i.e. $T_m (P)$) is an analytic function and well described by a smooth parameterized form. This assumption holds true unless the system has a phase transition along the melting line, i.e. a triple point. By examining the pair correlation function in fig~\ref{fig:gofr_triplepoint}, we observe a molecular-atomic transition in the neighborhood of 190 GPa for the classical system and 172 GPa for quantum hydrogen. Hence, we limit our calculation of the melting curve of classical hydrogen to the pressure range 50 GPa $\leq$ P $\leq$ 180 GPa and quantum hydrogen to 50 GPa $\leq$ P $\leq$ 170 GPa.
Within this range, we assume that the melting temperature can be described by a polynomial: $T_m (P) = \sum_{k=0}^M a_k P^k$ where $M$ is the polynomial order. While  other analytic forms, such as a Kechin form\cite{Kechin2001}, could be used, we found a polynomial satisfactory. Our calculations of the pressure derivatives  shown in fig. \ref{fig:melt} reveal a linear dependence on the pressure in this range, implying that a quadratic form is sufficient to fit the melting curves.

The CC relation only establishes the slope of the melting curve. We use the TP method described earlier to establish its absolute value. We construct a probability density to quantify how likely a proposed melting curve describes the data that we have---the CC values and the TP bounds:
\begin{equation}
\begin{split}
    \Pi({\bf a})= \exp \Bigg( -\chi^2 /2 
    -\sum_k  f(T_L(P_k)-T_m (P_k;{\bf a} )) \\
    -\sum_k  f(T_m(P_k;{\bf a} )-T_U (P_k)) \Bigg)
    \label{eq:prob}
\end{split}
\end{equation}
where $ \{T_L ,T_U \}$ are the lower and upper bounds at pressures $P_k$, $f(x)=\ln(1+\exp(x/\Delta T))$ is a function which constrains bound violations to be on the order of the parameter $\Delta T$ (typically 0.1 K) and  $\chi^2$ is the weighted squared deviation of the slope from the CC values:
\begin{equation}
    \chi^2 = \sum_k \frac{1}{\sigma^2 (P_k)}\left( D(T,P_k)-\frac{dT_m(P_k;{\bf a})}{dP} \right)^2.
\end{equation}
We have neglected multiplying by a
prior probability for the  coefficients $\bf a$. If these parameters are tightly constrained by the data, a variation in the prior will be unimportant. 

The probability distribution $\Pi({\bf a})$ is sampled using Markov Chain Monte Carlo. The sampling gives the most likely melting curve: the polynomial that minimizes $\chi^2$ while satisfying the bounds. The spectrum of sampled melting curves is used to estimate the overall uncertainty. We quote the interval that encompasses 90\% of the sampled curves. The sampling is extremely rapid: allowing millions of melting curves per second on a desktop computer. In addition, this scheme is quite flexible and can incorporate additional constraints, such as experimental data.

The error estimates using the combination of the TP method with the CC values are smaller than that provided by the bounds alone. We find uncertainties in the classical melting temperature to be about 5 K for pressures between 90 GPa $\leq$ P $\leq$ 130 GPa, much smaller than the TP bounds. Tight bounds at a single pressure, in combination with the CC estimates are sufficient to constrain the curve at all pressures. There are larger errors at the highest and lower pressures since there is less information to constrain the fit.


\subsection{Micro-canonical MD}

We have also used the micro-canonical method for estimating the melting temperature. A two phase system as described above is prepared for a given pressure and temperature near the melting transition using a barostat and thermostat. The thermostat and barostat are then removed, allowing the system to evolve under constant volume/energy conditions. After attaining equilibrium, the temperature is determined from the average kinetic energy, and the pressure from the virial.  

In the thermodynamic limit, this procedure will equilibrate precisely on the melting curve as long as the system ends up with substantial amounts of both phases. However, in practice, the results can depend on the extent of the simulation cell in the three directions, the preparation of the initial state, whether equilibrium is attained within the trajectory, and whether the solid is stress free.  In preparing the system, one does not necessarily have a solid with fully filled layers.  If the equilibrated state contains partially filled molecular crystal layers, this will increase the internal energy and hence decrease the observed temperature to be below the thermodynamic melting curve. 

Using this approach, we find results above or below the results from the TP-CC approach discussed above by about 30 K. The statistical errors in the temperature are about 15 K. Repeated runs targeted at the same pressure (150GPa) showed a spread in final temperatures of up to 40 K. Possibly a more careful preparation of the initial conditions combined with much longer MD trajectories could narrow the differences.

Applying the micro-canonical approach to quantum path integrals is more complicated, as the temperature also enters into the quantum action.  In contrast, the two methods we discussed above can be used for classical or quantum systems without change.

\subsection{The effect of quantum zero point motion}

The TP method was used to establish the quantum melting curve in Niu et al. \cite{Niu2023} and a combination of TP and CC was done for atomic hydrogen in Ly et al.\cite{ly2024}. However,  TP simulations require considerable computer resources, particularly to attain tight bounds and to account for the quantum effects of the protons. Path Integral methods are slower than classical simulations because of the extra degrees of freedom of the imaginary time path.
To estimate the melting temperature for quantum hydrogen, we start from the melting curve of the classical system and utilize an analogous method to the CC relation for pressure, previously employed to find the isotopic dependence of the melting line of solid helium at high pressure\cite{Boninsegni1994}, the kinetic energy of lithium\cite{Filippi1998} and the melting of atomic hydrogen at very high pressure\cite{Libertore2011}. 

We can determine how the melting pressure $P(T_m,x)$ at a temperature $T_m$ changes with the mass of the nucleus $m$, (written in terms of $x=m_p/m$) by differentiating the Gibbs free energy with respect to  $x$. For a point on the melting curve:
\begin{equation}
 \frac{dP(T_m,x)}{dx} =- \frac{\Delta k(P,T_m,x)}{x\Delta v(P,T_m,x)}
\label{eq:cons_T}
\end{equation}
where $\Delta v(P,T_m,x)=v_l(P,T_m,x)-v_s(P,T_m,x)$ is the volume difference per atom between the liquid and solid phase
and $\Delta k$ the similar difference in the kinetic energy per atom. 
To use this expression, we perform PIMD simulations on homogeneous liquids and solids for values of $(P,T_m,x)$ for 1/3 $\leq$ x $\leq$ 1, and pressures around the suspected melting temperature to determine the differences in kinetic energies and volumes. 
We expect that the melting pressure, $P_m(T_m,x)$, is a smooth function of x, assuming no phase transition occurs, so we expand $P_m(T_m,x)= P_c + P_1 x + \ldots$ where $P_c(T_m)$ is the classical transition pressure.
The expansion coefficients are determined by minimizing the squared difference between the fitted pressure derivative and the right hand side of Eq.~\ref{eq:cons_T} at fixed values of $x$  and interpolating in pressure. 
We performed this fitting at three temperatures (1142K, 1342K and 1500K) where it was determined that classical hydrogen melts at 72 GPa, 100GPa and 125GPa respectively. By using quantum systems of 192 and 360 protons and values of inverse mass $x= \{ 0.33, 0.5, 1.0 \}$ we determined the numerator and denominator of  Eq. \ref{eq:cons_T} to estimate the shift in melting temperature caused by the additional quantum fluctuations of the ions.

Once a point on the quantum melting curve is established, the CC method described earlier can be used to extend it to higher and lower pressures.  Values of the derivative of the quantum melting point as a function of pressure are shown in Fig.~\ref{fig:melt}. As with the classical system, the quantum derivatives follow a nearly linear relation in pressure, although a better fit was given by a cubic polynomial. 

The advantage of this procedure is that one can use the less expensive classical simulations to find a reference point on the classical melting curve. The change induced by the proton zero point motion is expressed in terms of the difference in the kinetic energies of the liquid and solid and determined by homogeneous simulations, not the more difficult and costly two phase procedure. One can obtain the melting curve for both hydrogen and deuterium at the same time.

We do not take into account the spin of the protons when performing the PIMD simulations;  we use distinguishable-particle path integrals. This corresponds to ``normal'' hydrogen, a mixture of 25\% para- and 75\%  ortho-hydrogen. Although DAC experiments typically load their samples at low temperatures where the hydrogen could have been predominately para-hydrogen, it is likely that it would have converted to normal as the temperature was raised. The effect of quantum statistics can be treated with PIMC as was done with superfluid helium\cite{ceperley1996} but it is not important for the melting line of molecular hydrogen at high pressure.

\section{Results}

\begin{figure}
    \centering
    \includegraphics[width=1.0\linewidth]{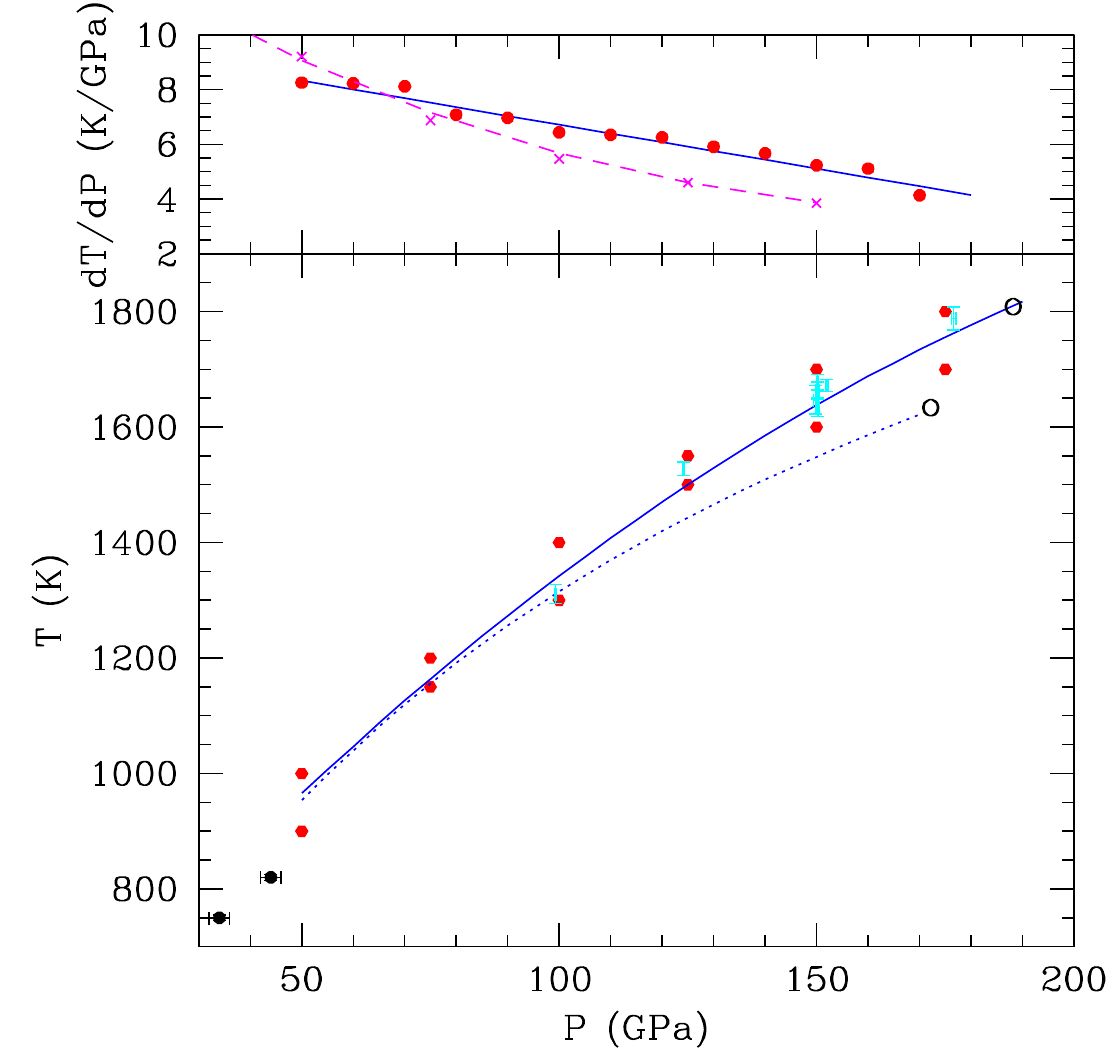}
    \caption{ Melting temperature versus pressure (lower panel),  the derivative of melting temperature with respect to pressure (upper panel). Blue lines are the best polynomial (lower panel) and its derivative (upper panel). Solid blue lines are classical hydrogen, dotted magenta lines, quantum hydrogen. The red points are simulation data used to determine the fit: upper and lower bound estimates in the lower panel and  CC estimates of the derivative in the upper panel. Magenta crosses in upper panel are the quantum estimates. The cyan points in the lower panel are estimates from microcanonical simulations. The black symbols are experimental estimates\cite{Gregoryanz2003} of the hydrogen melting transition. The two open circles on the melting curves are estimates of the molecular-atomic cross-overs for classical and quantum hydrogen. }
    \label{fig:melt}
\end{figure}

Tables \ref{table1}-\ref{table2} and Fig. \ref{fig:melt} present the estimated melting temperatures of classical and quantum molecular hydrogen for pressures between 50 GPa and 180 GPa. $T_m$(P) is the polynomial that optimizes the probability in Eq. (\ref{eq:prob}.), constrained to have a derivative consistent with the CC data (shown in Fig. \ref{fig:melt}) and to obey the TP bounds.
Tables \ref{table1}-\ref{table2} have the latent heat, internal energy of the solid, and volumes of the 2 phases along the estimated melting curve. These values were used to determine the CC derivative in Eq. (\ref{eq:D}).
We find that the classical melting temperature (in K) is reasonably well described by a quadratic function of pressure in the range 50 GPa $\leq$ P $\leq$ 170 GPa with polynomial coefficients  \{509,9.935,-0.01606\} in powers of pressure  (in GPa). The quantum (hydrogen melting line) is described by a quartic polynomial:  
        \{374,14.51,-0.0661,1.67E-4,-1.54E-7\} over the same pressure range.
We note a larger slope of the quantum melting curve below 50 GPa seen in the simulations and also in the experiments\cite{Datchi2000}. 
We performed simulations at 25 GPa with the M18 model, but we did not include these data in the fit since their densities are outside the range of the training data set.

The tables show that the liquid volume is about 1.4\% larger than the solid volume for all pressures in both classical and quantum hydrogen within this pressure range. This relative volume difference is half the difference calculated with the Silvera-Goldman pair potential\cite{Silvera1978} at 0.6 GPa \cite{Young1981}. 
We also find that the internal energy of the solid is higher than that of the liquid. From the CC relation these results implies that the melting point maximum does not occur for P$\le$180GPa in either classical or quantum hydrogen. We do observe a progressive decrease of the melting line slope with pressure. 

At slightly higher pressure we observe a rapid change in the liquid phase from a predominantly molecular hydrogen to a predominantly atomic hydrogen as shown in the pair correlation functions in Fig.  \ref{fig:gofr_triplepoint}. This occurs at roughly 189 GPa in classical hydrogen and at roughly 172 GPa in quantum hydrogen. These values are also shown on the phase diagram in Fig. \ref{fig:melt}. We find that the character of the melting line changes drastically above these pressures.  These pressures are close to those found in CEIMC simulations examining the liquid-liquid phase transition\cite{Pierleoni2016b}.   We plan future work looking at the nature of this atomic-molecular transition and its effect on the melting curve. More training data that targets dense hydrogen at pressures above 200 GPa and a different ML model are needed to make confident predictions about the continuation of the melting line above the molecular-atomic crossover.

\begin{figure}[h]
\centering
\includegraphics[width=\linewidth]{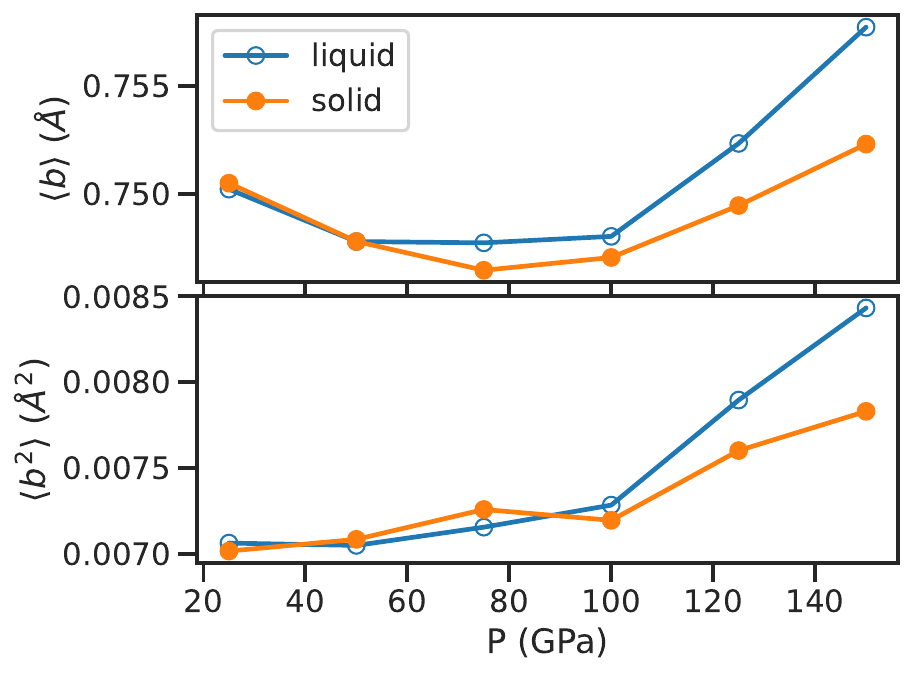}
\caption{Average bond length and variance, in \AA~and \AA$^2$, of quantum molecular hydrogen using the M18 model along the melting curve.}
\label{fig:bond-dist}
\end{figure}

By examining the distribution of molecular bond lengths, we find a change in the molecules in the liquid at pressures higher than $100$ GPa. As shown in Fig.~\ref{fig:bond-dist}, both the bond length and its fluctuation increase significantly from P=$100$ to $150$ GPa. This could be a precursor of the atomic-molecular transition at 170 GPa. The solid phase shows a smaller change.
The vibron frequency should show a change above 100 GPa since it is inversely proportional to $<b^2>$ in the harmonic approximation.
Table \ref{table4} reports the excess kinetic energy, i.e. the kinetic energy above the classical kinetic energy, of the solid along the melting line. This excess is primarily coming from the energy of the vibron and should approach the value for an isolated molecule at low density and temperature. The liquid has a slightly lower excess kinetic energy than the solid but it is only 2\% lower at 131 GPa. 

\begin{table}
\begin{tabular}{|l|l|l|l|l|l|}
\hline
P (GPa)	&T$_m$ (K)	&L  	&v$_s$ 	&v$_l$ 	& u$_s$ \\
\hline
50	&966	&6.6		&3.916	&3.973  & -2970.6\\
60	&1047	&6.4		&3.671	&3.710  & -2954.7\\
70	&1126	&7.8	    &3.444	&3.500  & -2939.0\\
80	&1201	&8.8		&3.271	&3.318  & -2924.6\\
90	&1273	&8.4		&3.123	&3.165  & -2909.5\\
100	&1342	&6.3		&2.996	&3.025  & -2895.0\\
110	&1408	&9.7		&2.876	&2.920  & -2881.6\\
120	&1470	&8.2		&2.779	&2.814  & -2868.4\\
130	&1529	&10.5       &2.689  &2.731  & -2855.6\\
140	&1585	&11.0		&2.611	&2.651  & -2843.3\\
150	&1638	&11.5		&2.541	&2.577  & -2831.4\\
160	&1688	&12.0		&2.473	&2.509  & -2819.5\\
170	&1734	&13.0		&2.412	&2.443  & -2808.1\\
180	&1777	&15.9		&2.356	&2.369  & -2797.4\\    
\hline
\end{tabular}
\caption{Melting line properties of classical molecular hydrogen  using the M18  model. The latent heat (L) and internal energy of the solid $u_s$ are in KJ per mole of H$_2$. The volumes are in cc/mole. 
The uncertainty in the melting temperature, estimated from the confidence intervals of the MCMC simulation, is about 5K. 
The errors in the latent heat  vary from 0.2KJ/mol for P<100GPa  to 1KJ/mol at 170GPa. The errors in the volumes are  0.001 cc/mol. The error  in internal energy of the solid is less than 0.1KJ/mol.
}
\label{table1}
\end{table}

\begin{table}
\begin{tabular}{|l|l|l|l|l|l|}
\hline
P (GPa)	&T$_m$ (K)	&L  	&v$_s$  	&v$_l$ & u$_s$ 	\\
\hline       
50	&954	&4.3		&3.944	&3.986  & -2950.0\\
75	&1156	&7.4	    &3.371	&3.415  & -2912.8\\
100	&1315	&7.9		&3.003	&3.036  & -2877.7\\
125	&1443	&9.0		&2.742	&2.770  & -2846.7\\
150	&1548	&10.0		&2.544	&2.569  & -2818.0\\
\hline
\end{tabular}
\caption{Melting line properties of quantum molecular hydrogen  using the M18  model. For units and errors, see caption for Table \ref{table1}.}
\label{table2}
\end{table}

\begin{table}
\begin{tabular}{|l|l|l|l|l|l|l|}
\hline
P (GPa)	&T (K)	&$k_{ex} (K)$  	&$\Delta k$ (K) &dP(1)/dx & dP(2)/dx &dP(3)/dx	\\
\hline
51   & 960  & 627      &  5  &  0.9  & 0.6  & 0.5 \\
73	&1140	& 600	    & 7	& 2.5 & 1.0 & 0.9\\
102	&1340	&	561	&	6 & 2.5 & 1.7 & 2.0 \\
131	&1500	&	525	& 10 	& 4.4 & 2.0  & 3.9 \\
\hline
\end{tabular}
\caption{ The data used to compute the change in melting pressure due to quantum effects of the ions.  P and T are the pressure and temperatures where the derivatives were estimated.
k$_{ex}$ is the excess kinetic energy in the solid phase, $\Delta k$ is the difference in kinetic energy between the solid and liquid, both in K per ion for $x=1$ and have errors of 1K. dP (m/m$_p$)/dx are the estimated derivatives in GPa from Eq. \ref{eq:cons_T} for mass m. Errors on the derivatives are 0.5 GPa.
}
\label{table4}
\end{table}

\begin{figure}
    \centering
    \includegraphics[width=0.97\linewidth]{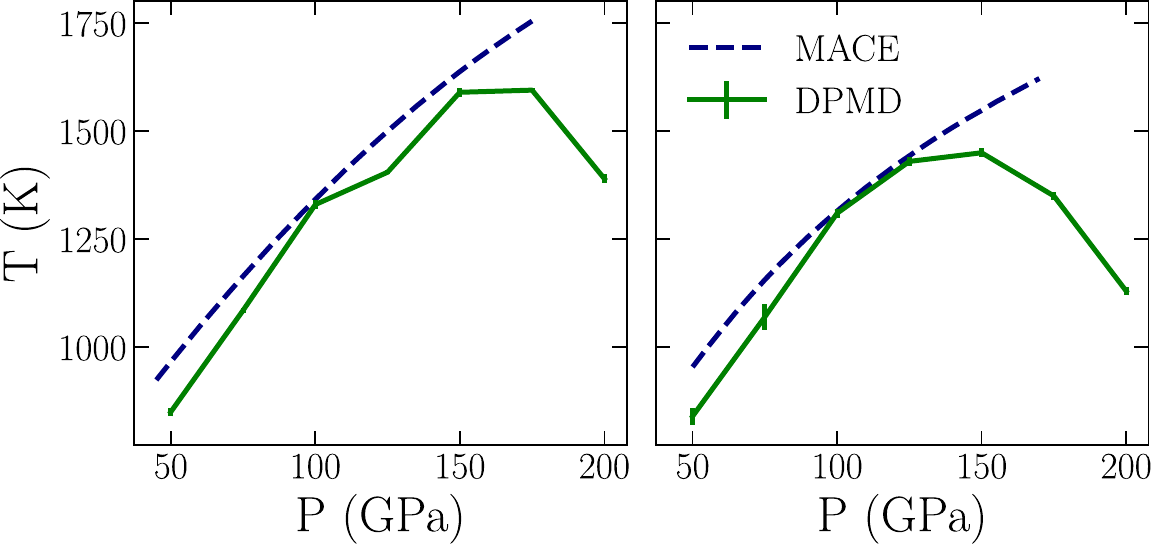}
    \caption{Comparison of melting lines derived from two ML models trained on QMC data: classical hydrogen (left) and quantum hydrogen (right). The MACE M18 model exhibits a higher melting line compared to DPMD. The DPMD melting line was determined using a two-phase method, with the line positioned midway between the upper and lower bounds with error bars indicating those bounds.}
    \label{fig:M18_vs_DPMD_melting}
\end{figure}

\section{Discussion}

For classical hydrogen, the M18 model predicts a higher melting temperature than that of the previous DPMD model\cite{Niu2023} trained on the same data (Fig.~\ref{fig:M18_vs_DPMD_melting}). The melting temperature from DPMD simulations using the TP method with classical protons agree with the prediction of the M18 model at 100 GPa. However, at other pressures, the DPMD melting temperatures are 50-100 K lower, except at 175 GPa where it is 160 K lower.  We attribute some of this spread in difference in melting temperatures to ambiguities in using the TP method discussed above, as highlighted by the dip in the classical DPMD model at 125 GPa. The emphasis on weighting the energy in training the M18 model may have resulted in an increased stability of the classical solid. 
Remarkably, the melting temperatures of quantum hydrogen predicted by the M18 model are consistent with those of the DPMD model except at 150 GPa where the DPMD model predicts melting at 80 K lower temperature.  
This differene between the two models at 150GPa is possibly related to the nearby atomic transition in the quantum liquid phase at 172 GPa and at 190 GPa in the classical liquid phase. Notably, our study of the melting curve does not include atomic liquid configurations.

Using the M18 model, we did not see the solid-solid phase transitions from hcp to Fmmm-4 reported by the DPMD model trained on the same QMC data~\cite{Niu2023}. 
Our CC and TP runs in the solid phase employed flexible boxes, allowing for the system to dynamically change crystal structure, although dynamical barriers may exist.
This absence of the solid-solid transition likely stems from the differences in the models, warranting further investigations.
It is worth noting that static experiments did not report any solid structure changes in the Raman spectra along the melting line for  P $\leq $ 70GPa \cite{Gregoryanz2003}.
Overall, despite differing optimization, weighting, and architectures, the two ML models trained on the same dataset and analyzed using distinct methodologies reach a consistent conclusion: solid molecular hydrogen remains stable at temperatures above 1600 K. 



Experimental determination of the melting line become increasingly difficult as pressure increases. Below roughly 40 GPa, static compression methods can directly observe a physical change of melting, for example by optically detecting the presence of a mixed solid-liquid sample. At higher pressures, however, the extreme conditions required often damage/destroy the diamond anvil containing the sample, necessitating dynamic experiments. These involve localized heating of the sample for a brief period of time during which optical measurements are made.
At lower pressures, melting temperatures are determined through the difference between the index of refraction of the liquid and crystal up to 8 GPa by Diatshenko et al. \cite{Diatschenko1985} and up to 15.2 GPa by Diatchi et al.\cite{Datchi2000}. Gregoryanz at al. \cite{Gregoryanz2003} extended these melting line measurements from 10 GPa to 45 GPa using Raman spectroscopy, observing a positive discontinuity in the molecular vibron peaks. However, the method's sensitivity diminishes with increasing pressure, and no clear transition was identified at higher pressures for temperatures up to 1000 K.
Their melting temperatures are shown in Fig. (\ref{fig:melt}).  The estimations of melting from the M18 model are consistent with the measurements of Gregoryanz et al.'s results. Although we estimate melting temperatures at lower pressures, and they are also in agreement, we do not report them here, as such pressures are outside of the range where we have training data.

At higher pressures, laser heating must be employed to avoid breaking the diamonds resulting in much more uncertain results. Furthermore, the composition of the sample, its temperature and pressure are very uncertain. These conditions often deviate from unstrained, equilibrium state of pure hydrogen, complicating the interpretation of results.  In addition,
the measurements that report melting, e.g. using Raman spectroscopy, do not provide direct 
evidence of the presence of liquid or crystal phases.  

The highest experimental melting temperature, reported by Deeymad and Silvera \cite{Deemyad2008} utilized pulsed laser heating to observe melting via laser speckles and a temperature plateau with increased laser heating. These results agree with the lower-pressure measurements of Gregoryanz et al. \cite{Gregoryanz2003}. 
Subsequent measurements by Eremets and Troyan \cite{Eremets2009} suggest a much broader maximum
near 100 GPa based on changes in the laser speckle patterns and a large resistivity drop. 
Subramanian et al. \cite{Subramanian2011} extended Raman measurements of the vibron up to 140 GPa. To identify melting at higher pressures Subramanian et al.~\cite{Subramanian2011} took into account thermal gradients, the onset of speckle motion and the lower frequency vibron in the fitting spectrum above 30 GPa. 

Howie et al. \cite{Howie2015} traced the presumed melting transition up to 250 GPa by observing a discontinuity in the vibron shift. Similarly, Zha et al. \cite{Zha2017} determined melting from 140 GPa to 300 GPa, identifying a melting point maximum of 825 K at 70 GPa 
based on the disappearance of low-frequency lattice modes.
However, the existence of thermal gradients within the sample complicates the estimation of the temperature, suggesting these measurements represent a lower bound for the melting temperature.
Notably, no experimental data directly probe the pressure range where we predict the melting point maximum (172 GPa, 1630 K). Dynamic experiments conducted at pressures above 50 GPa have several questionable aspects and are not consistent in their results. See the discussion in the review by Goncharov et al. \cite{Goncharov2013}.


The interpretation of melting experiments has been supported by two theoretical ideas. First is the melting curve formula of Kechin\cite{Kechin2001}, explicitly constructed to have a melting point maximum. Datchi et. al\cite{Datchi2000} fit their data at pressures up to 15.2 GPa to a Kechin curve to obtain a maximum of 1100 K at 128 GPa. Meanwhile, Gregoryanz et al. \cite{Gregoryanz2003} find a lower maximum of 950 K and 77 GPa when they fit their data to a Kechin curve at pressures ranging from 7 GPa up to 45 GPa. However, a Kraut-Kennedy\cite{KrautKennedy} model, which fits equally well to the Gregoryanz data\cite{Gregoryanz2003}  does not predict a maximum.  As we commented earlier, our explicit calculations of the derivatives of the melting curve do not support having a maximum of the melting curve before the molecular-atomic crossover in the liquid phase at roughly 173 GPa.

The second theoretical influence comes from MD simulations using a PBE exchange correlation functional\cite{PBE} that find a melting point maximum. While the melting curves from various studies show relative consistency among each other, their maxima differ by several hundred kelvin. A detailed comparison is shown in the supplementary materials of Niu et al.\cite{Niu2023}.  
The differences between the various MD simulations arise from variation in system sizes, the bases for wavefunction expansion, the methods for determining melting and other technical details, highlighting the sensitivity to methodological choices. 
For instance, Bonev et al.~\cite{Bonev2004} performed AIMD simulations with 720 atoms using the TP method for melting, estimating a maximum at 820 K and 82 GPa.
Morales et al.\cite{Morales2010} verified these results using a free energy method with 400 classical protons obtaining a maximum of 925 K at 106 GPa.
Similarly, Liu et al.\cite{Liu2013}  performed a TP simulation with 1960 atoms in the microcanonical ensemble and found a maximum melting temperature of 1150 K at 100 GPa.
Zong et al. \cite{Zong2020} used a ML potential trained on PBE forces to simulate dense hydrogen using the z-method with up to 70,000 atoms and found a maximum melting temperature at 900 K at 90 GPa. Cheng et al.\cite{Cheng2020} trained a Behler-Parrinello model\cite{Behler2007} across a wide range of pressures including both molecular and atomic hydrogen using two different functionals, PBE\cite{PBE} and BLYP\cite{Becke1988}, and examined melting via quenching or superheating of cells up to 1728 atoms. Using the PBE\cite{PBE}  functional they reported a maximum melting temperature at 850 K $\pm$ 200 K at 125 GPa while the BLYP\cite{Becke1988} functional resulted in a maximum temperature of 1000 K $\pm$ 150 K at 150GPa.  
Because the PBE functional does not have dispersion interactions, its results favor the liquid phase over the solid phase. Nonetheless, the benchmarking studies comparing QMC energies with those of a functional are a good indication of how accurate a functional will be on the melting curve. For example, the vdW-DF1 functional would be a better choice than PBE as evidenced by comparisons with the DPMD melting curve\cite{Niu2023}, the equation of state and optical properties near the atomic-molecular transition\cite{Gorelov2019} and the direct comparison with QMC energies and forces on configurations\cite{Clay2016,Ceperley_2024} as summarized in Table \ref{tab:hydrogen errors}. However, while the vdW-DF1 functional has a larger rms error of 124 meV compared to the QMC energies, the M18 model achieves a lower RMS error of 29 meV.
The AIMD and ML studies, suggest that an accurate ML model can reproduce direct AIMD simulations of the melting curve. However, one can use the ML model on larger systems, both in space and time. If it is trained on QMC data it will be more predictive.

\section{Summary and Outlook}

We have shown that the finding of a higher melting temperature for dense molecular hydrogen is confirmed by using a more accurate ML model specifically trained to have smaller energy errors.  The accuracy of the model is established/benchmarked by comparing to experimental scattering data, by observations for pressures at P $\leq$ 44GPa and by comparing directly to CEIMC simulations.  To confirm the calculated melting curve, we have used several different procedures for determining the melting temperature in the range of pressures from 50GPa to 180GPa including two phase simulations, estimations of the slope of the melting curve using the Claudius-Clapeyron relation, microcanonical simulations and using the change of melting curve with respect to the proton mass.

There are many additional questions about the phase diagram of molecular hydrogen that this ML model could address. In addition to the melting line, it is important to look at the low temperature solid-solid transitions in the molecular phase that have been seen experimentally. The current model should be adequate inside the pressure range 50GPa $\leq$ P $\leq$ 200GPa.  We plan on making a comprehensive equation of state model using the M18 model.
Within the solid phase, the M18 model can be used to determine the Raman spectra as measured in experiment, as well as other properties.
Of course further work can be done to improve the model, to determine the limitations of the QMC energies and forces, to refine the selection of training configurations and to explore other forms for the ML model, in particular to include long-range dispersion interactions.

We are currently investigating the details of the triple point on the melting curve and how that affects the molecular-atomic (i.e. the liquid-liquid) transition. 
Preliminary results show that the melting curve drops quickly after the cross-over. Tracing this melting line as the molecular crystal becomes metallic and then atomic will aid in interpreting experimental studies at higher pressures.  Molecular dynamics of realistic ML models in combination with the tools of path integrals, constant pressure and temperatures can finally resolve the phase diagram of the first element. 

\section{Acknowledgement}

The Flatiron Institute is a division of the Simons Foundation. D.M.C., S.G. and S.J. are supported by DOE DE-SC0020177. 
CP was supported by the European Union - NextGenerationEU under the Italian Ministry of University and Research (MUR) projects PRIN2022-2022NRBLPT CUP E53D23001790006 and PRIN2022-P2022MC742PNRR, CUP E53D23018440001.
We thank the QMC-HAMM team and H. Niu for valuable discussions. 
Comparison of melting lines from various calculations were performed on
the Illinois Campus Cluster and Delta, supported by the National Science Foundation (Awards No. OCI-0725070 and No. ACI-1238993), the state of Illinois, the University of Illinois at Urbana-Champaign, and its National Center for Supercomputing Applications. 

\appendix

\section{Methods}
Density functional theory (DFT) and diffusion Monte Carlo (DMC) was used to evaluate the ground state forces experienced by a collection of $N$ hydrogen atoms in a supercell.
Given the proton coordinates $R_I,I=1,\dots, N$, the ground-state wavefunction $\Psi_0(\left\{\bs{r}_i,i=1,\dots,N\right\}|R)$ of the electrons is the lowest-energy eigenvector of the many-body Hamiltonian. 
Starting from a trial wavefunction $\Psi_T (r|R)$, DMC stochastically executes imaginary-time evolution via a drift-diffusion-branching process to sample the mixed distribution $f=\Psi_T^*\Psi_0$~\cite{Martin_Reining_Ceperley2016}.
Forces were computed using the Hellmann-Feynman theorem with regularization to control the variance of the estimator~\cite{Chiesa2005}.

DFT replaces the electron-electron interaction with a single-particle effective potential, turning the many-body Hamiltonian into a self-consistent mean-field one-body Hamiltonian.
The accuracy of DFT is determined by the choice of density functional. Common density functionals are 
the generalized gradient approximation (GGA), such as PBE~\cite{PBE}, and the van der Waals (vdW) corrected variants\cite{vdw-df}.
The choice of density functionals have been shown to greatly affect the calculated properties of dense hydrogen~\cite{Morales2013a}.
In contrast, DMC energies and forces have a much more reliable accuracy and have been used to benchmark DFT forces\cite{Clay2016} and to predict the first-order liquid-liquid transition~\cite{Pierleoni2016b}, afterwards verified by experiment~\cite{Celliers2018a}.

The mean of the DMC energy estimator is an upper bound to the exact ground-state energy.
Minimization of the energy and variance of the trial wavefunction allows one to systematically approach the exact ground-state~\cite{Martin_Reining_Ceperley2016}. While sharing the same $O(N^3)$ scaling as DFT,
DMC has a greater computational cost than DFT because of the explicit treatment of many-body correlations.
DFT calculations are important to build the single electron part (Slater determinant) of the DMC trial wavefunction.
The uncertainty on how to compute forces in QMC have hindered the use of DMC forces for research in dense hydrogen, for example to build empirical potential energy models.
The extensive database of DMC forces we calculated alleviates this problem.

The database contains a set of configurations and associated properties.
Each file contains a configuration (the three-dimensional positions of $N$ protons) and some properties with key metadata. Properties include
the simulation cell, a $3\times3$ real matrix, where each row is a lattice vector of the supercell and 
the energy and forces with their statistical QMC errors.
For some configurations, the stress tensor of the super cell is stored as a $3\times3$ real matrix.
The metadata includes the code version used to perform the calculation, the input to the code, and other useful information for reproducing the calculation.\footnote{The data and code can be found in the following url https://qmc-hamm.hub.yt/index.html}

\subsection*{Configuration Generation and Selection}

We used a variety of methods to generate configurations, then from those selected, a subset  was chosen for property evaluation. 
To create an empirical interatomic model potentials, it is useful to estimate the thermodynamic condition that the model will be applied to, e.g. the range of temperatures and densities.
This is especially important for deep neural networks,
which excel at interpolating non-linear functional forms in high-dimensional space, but have no guarantee of extrapolating to accurate results at conditions which do not have nearby training data.

Unless the model has correct asymptotic limits built in, the neural network can produce unphysical results if extrapolated too far from the training data.

In constructing the database, we focused on the thermodynamic conditions of melting of dense molecular hydrogen as shown in Niu et al. \cite{Niu2023}.
We use a variety of classical and quantum MD and MC methods to generate configurations
including Born-Oppenheimer path integral Monte Carlo using the CEIMC method \cite{Pierleoni2006}, DFT molecular dynamics (DFT-MD), classical molecular dynamics, and path-integral (PI) molecular dynamics (MD). 
We employed Quantum Espresso (QE)~\cite{Giannozzi2009,Enkovaara2017}, LAMMPS~\cite{LAMMPS}, and i-PI~\cite{i-PI2014}, respectively.
QMC calculations were carried out using the QMCPACK software ~\cite{Kim2018,Kent2020} and BOPIMC, an in-house code implementing CEIMC for classical and quantum protons. Configurations had 96 protons in an orthorhomic supercell.

To select a smaller subset of configurations for QMC calculations, we use a sparsification method to spread out selected configurations over a space of descriptors
chosen to characterize a configuration. These included the volume of the cell, energy of the configuration, the mean bond length for molecules. 
Given the chosen descriptors, the total configurations set, and the number of subset configurations needed, we implemented a simulated annealing approach with a cost function designed to spread the selected configurations maximally. 

\subsubsection*{QMC calculations of energy and forces}

A diffusion Monte Carlo (DMC) algorithm as implemented in QMCPACK  with a fixed number of electrons, twist-averaged boundary conditions~\cite{Lin2001} and a Slater-Jastrow (SJ) trial wavefunction,
\begin{equation}
\label{eq:sjwf}
\Psi(\bs{r}|R) = \text{det}(\{\phi_i(\bs{r}_j|R)\})  \exp\left(-U(\bs{r}|R)\right)
\end{equation}
is a product of determinants of single-particle orbitals and a Jastrow function $\text{exp}(-U)$ was used.
We use the occupied Kohn-Sham orbitals from a PBE calculation performed on a $4^3$ shifted uniform grid of k-points of the supercell. 
The DFT calculation used a hard pseudopotential with a core radius of $0.37$ bohr, generated using Opium~\cite{Grinberg2000}.
The Jastrow part contains electron-ion and electron-electron terms:
\begin{equation}
\label{eq:ujas}
U =  \sum\limits_{i,J} u_1(\bs{r}_i-\bs{R}_J) +  \sum\limits_{i<j} u_2(\bs{r}_i-\bs{r}_j).
\end{equation}
The electron-electron term is a sum of long- and short-range pieces.
The optimized wavefunction parameters are stored in the metadata of each data file.
Each DMC calculation performs $102.4$ ha$^{-1}$ of imaginary time propagation using a timestep of $0.02$ ha$^{-1}$ and a walker population of $2058$.

The mean and standard error of each property is calculated after discarding an equilibration period of $25.6$ ha$^{-1}$. The electronic density $\rho_{\bs{k}}$, structure factor $S_{\bs{k}}$, and forces are recorded in addition to the total energy. All properties are averaged over the $4\times4\times4$ shifted uniform twist grid and all, except the total energy, are linearly extrapolated (2*DMC-VMC) to reduce the mixed-estimator bias.
The typical standard errors of the energy/atom and force components were 0.1 meV and 130 meV/\AA \ respectively.
The processed fluctuating structure factor is interpolated to estimate the finite-size correction of the total energy, assuming an RPA form for the long-range part of the Jastrow pair function $u_{\bs{k}}$~\cite{Holzmann2016}.
No corrections are applied to the forces. 

We choose an energy cutoff of $46$ Rydberg, a $4^3$ shifted uniform k-grid for the DFT calculations and
 an energy cutoff of $200$ Rydberg to generate orbitals for QMC, since the pseudopotential has a small core cutoff of $0.37$ bohr.
All spline parameters in the electron-electron and electron-ion Jastrows are variationally optimized for each configuration to convergence. 
 We find similar energy variances for different configurations, which indicate consistent wavefunction quality.

The Hellmann-Feynman forces are expectation values of the derivative of the Hamiltonian with respect to proton coordinates
\begin{equation} \label{eq:hf-forces}
F_{I\alpha} = \langle \Psi_0 \vert \dfrac{\partial H}{\partial R_{I\alpha}} \vert\Psi_0\rangle,
\end{equation}
where $\Psi_0$ is the ground state of the Hamiltonian at fixed atomic coordinates $R_{I\alpha}$ ($\alpha=x, y, z$).
The Monte Carlo estimator is regularized\cite{Chiesa2005} to have finite variance.

DMC calculation of properties that do not commute with the Hamiltonian suffer from the mixed-estimator bias, because they are sampled from the mixed distribution $\Psi_T^*\Psi_0$ rather than the ground-state distribution $\Psi_0^*\Psi_0$. 
The mixed estimator bias is relatively independent of pressure but increases roughly linearly with temperature from $300$ meV/\AA~at $600$ K to $550$ meV/\AA~at $2200$ K. We apply a linear mixed-estimator correction to obtain QMC forces; the remaining bias is expected to be an order of magnitude smaller.
We apply structure-factor-based finite-size correction (FSC) to the QMC total energy~\cite{Holzmann2016}.
The energy per particle of the $N=96$ and $N=128$ calculations agree with each other within error bars after FSC at three tested densities from configurations generated at $1500$ K.

\subsubsection*{Simulation Details}

In this work, we trained MACE models with version 0.2.0 
with increasing complexity of architectures and larger atomic environment cutoff radii until the errors saturated. These studies were done keeping in mind the MD run times of a model. In the new model we had the hidden irreps parameter set to "128x0e + 128x1o" and a cutoff radius of 4.0-4.4 \AA. We then vary the energy and force weights to obtain the final model (M18) using $\lambda_E/\lambda_F = 220$ \AA$^{-2}$. 

The classical MD simulations were performed using the LAMMPS software\cite{LAMMPS}  with a time step of 0.5fs. The quantum PIMD simulations were performed using i-PI\cite{i-PI2014} with a time step of 0.1 fs.  The number of  path integral beads was approximately  10$^4$/T where T is the temperature in K.   For the CC runs to compute the energy and volume we used 360 atoms in the classical simulations and 192 atoms for quantum hydrogen. For the TP classical runs we used 3072 atoms.  In all cases we used thermostats and barostats. For the liquid CC runs we assumed a cubic box, while for the solid CC runs we had a flexible box.


\bibliography{ref}
\end{document}